\documentclass[aps,prd,twocolumn,groupedaddress,showpacs,nofootinbib]{revtex4}
\usepackage{graphicx,amssymb}

\begin{document}

\title{PPN-limit of Fourth Order Gravity inspired by Scalar-Tensor Gravity}

\author{S. Capozziello$^{\dagger}$,
A. Troisi $^{\diamond}$}
\thanks{Corresponding author\,: A. Troisi, {\tt antro@sa.infn.it}}
\affiliation{$^\dagger$ Dipartimento di Scienze Fisiche, Universit\`{a} di Napoli ``Federico II", INFN, Sez. di
Napoli, Compl. Univ. di Monte S.Angelo, Edificio G, Via Cinthia, I\,-\,80126\,-\, Napoli, Italy
 \\
$^\diamond$  Dipartimento di Fisica ``E.R. Caianiello", Universit\`{a} di Salerno,
 INFN, Sez. di Napoli, Gruppo
Collegato di Salerno,
 Via S. Allende, I\,-\,84081\,-\,Baronissi (SA), Italy}

\begin{abstract}
Based on the {\it dynamical} equivalence between higher order gravity and scalar-tensor gravity the PPN-limit of
fourth order gravity is discussed. We exploit this analogy developing a fourth order gravity version of the
Eddington PPN-parameters. As a result, Solar System experiments can be reconciled with higher order gravity, if
physical constraints descending from experiments are fulfilled.
\end{abstract}


\maketitle






\section{Introduction}

The recent debate about the origin of the cosmic acceleration, induced by the results of several astrophysical
observations \cite{sneia,cmbr,lss}, led to investigate several theoretical approaches capable of providing viable
physical mechanisms to the dark energy problem. In this wide discussion, no scheme seems, up to now, to furnish a
final answer to this puzzling conundrum. Nevertheless among the different models, ranging from quintessential
scenarios \cite{steinhardt}, which generalize the cosmological constant approach \cite{starobinsky-shani}, to
higher dimensional scenarios \cite{braneworlds,dgp} or the resort to cosmological fluids with exotic equation of
state \cite{chaplygin,vdw} and unified approaches considering even dark matter \cite{hobbit,unified}, an
interesting scheme which seems to deserve a major attention is represented by higher order theories of gravity.
This approach obtained by the generalization of the Einstein gravity, has led to interesting results both in the
metric formulation \cite{curv-quint,noi-ijmpd,noi-review,carroll,odintsov-m} and in the Palatini one
\cite{palatini,francaviglia}.\\
Recently some authors have analyzed the PPN-limit of such theories both in the metric and in the Palatini approach
\cite{olmo,allemandi-ruggiero} with contrasting results. Thus, it seems interesting to deepen the discussion about
the Post Parametrized Newtonian (PPN) behaviour of this theory. The purpose is to verify if the cosmological
reliability of such a scheme can be drawn even on the Solar System scales and to understand if the hypothesis of a
unique fluid working as a two ``faces" component (matter and geometry) can be a workable one.
\\
In this paper we exploit the strict analogy between the higher order gravity and the scalar-tensor theories to
develop a PPN-formalism for a general fourth order gravity model in the metric framework, working, in general, for
extended theories of gravity. There are strong analogies between these two approaches. The similarity between the
non-minimally coupled scalar models and the higher order gravity ones is known since 1983 \cite{teyssandier83},
when it was demonstrated the similarity between a scalar-tensor Lagrangian of Brans-Dicke type and fourth order
gravity. Actually, such an interpretation goes well beyond conformal transformations, since it is a formal analogy
without any physical change in the dynamical
variables of the system.\\
In this paper, we further discuss the analogy between fourth order gravity and scalar-tensor gravity considering
the PPN-parametrization descending from such a similarity. As main result, we show, despite some recent studies
\cite{olmo}, that Solar System experiments do not exclude the possibility that higher order gravity theories can
represent a viable approach even at scales shorter than the cosmological ones. In other words, standard General
Relativity should be revised both at cosmological and Solar System distances in order to solve several mismatches
between the theoretical predictions and the observational results.

\section{Fourth Order Gravity vs. Scalar Tensor Gravity}

Let us recall how the analogies between the two schemes arise. As it is well known, scalar-tensor gravity is
obtained if a scalar-field-matter Lagrangian is non\,-\,minimally coupled with the Hilbert-Einstein Lagrangian.
The general action for a such theory is \cite{capoz-scal-tens}:
\begin{equation} \label{L-scal-tens}
{\cal{A}}=\int\,d^4x \sqrt{-g}\left[F(\phi)R+\frac{1}{2}g^{\mu\nu}\phi_{;\mu}\phi_{;\nu}-V(\phi)+\kappa {\cal
L}_m\right]\,,
\end{equation}
where $F(\phi)$ is the coupling function, $V(\phi)$ the self-interaction potential, $\phi$ a scalar field, ${\cal
L}_m$ the ordinary matter Lagrangian and $\kappa$ the dimensional coupling. This relation naturally provides
Brans-Dicke gravity \cite{brans-dicke} if it is rearranged through the substitutions\,: $\varphi\,=\,F(\phi)\,,
\quad \omega(\varphi)\,=\,-\displaystyle\frac{F(\phi)}{2F'(\phi)^2}$ \cite{capoz-brans}; its peculiarity is to
account for Mach principle which leads back inertial forces within the background of gravitational
interactions.\\
The $f(R)$ gravity action in the metric formalism is the following \cite{curv-quint,noi-review}
\begin{equation}\label{eq:def-f(R)}
{\cal {A}}=\int d^4x\ \sqrt{-g}\left[f(R)+\kappa {\cal {L}}_m\right]\,,
\end{equation}
which depends on the metric $g_{\mu \nu }$ and the matter fields. Again $\kappa$ defines the dimensional coupling.
The energy-momentum tensor of matter is given by the relation $\displaystyle{T^{m}_{\mu \nu
}=\frac{-2}{\sqrt{-g}}\frac{\delta {\cal L}_m}{\delta g^{\mu \nu }}}$.\\
From the action (\ref{eq:def-f(R)}), we obtain the fourth order field equations\,:
\begin{equation}\label{eq:f-var}
f^\prime(R)R_{\mu\nu}-\frac{1}{2}f(R)g_{\mu\nu}=f^\prime(R)^{;\alpha\beta}\left(g_{\mu\alpha}g_{\nu\beta}-
g_{\mu\alpha}g_{\alpha\beta}\right) +\kappa\,T^m_{\mu\nu}\,,
\end{equation}
which can be recast in a more expressive form as:
\begin{displaymath}
G_{\mu\nu}=
\frac{1}{f^\prime(R)}\bigg\{\frac{1}{2}g_{\mu\nu}\left[f(R)-f^\prime(R)R\right]+{f^\prime(R)}_{;\mu\nu}+ \bigg.
\end{displaymath}
\begin{equation}\label{eq:f-var2}
\bigg.-g_{\mu\nu}\Box{f^\prime(R)}\bigg\}+ \frac{\kappa}{f^\prime(R)}T^m_{\mu\nu}\,,
\end{equation}
where $G_{\mu\nu}$ is the Einstein tensor and $f'(R)\equiv df/dR$; the two terms ${f^\prime(R)}_{;\mu\nu}$ and
$\Box f'(R)$ imply fourth order derivatives of the metric $g_{\mu\nu}$. On the other side, if $f(R)$ is a linear
function of the scalar curvature, $f(R)=a+bR$, the field equations become the ordinary second-order
ones.\\
Considering the trace of Eq.(\ref{eq:f-var2}), 
\begin{equation}\label{eq:f-var-trace}
3\Box f'(R)+f'(R)R-2f(R)=\kappa T\,.
\end{equation}
Such an equation can be interpreted as the equation of motion of a self-interacting scalar field, where the
self-interaction potential role is played by the quantity $V(R)\,=\,f'(R)R-2f(R)$. This analogy can be developed
each time one considers an analytic function of $R$ which can be algebraically inverted so that $R$ reads as
$R=R(f')$, in other words it has to be $f''(R)\neq 0$. In fact, defining
\begin{eqnarray}\label{eq:phi=f'}
\phi&\equiv& f'(R)\\\nonumber \\V(\phi)&\equiv& R(\phi)f'(R)-f(\phi) \label{eq:V=rf'-f}
\end{eqnarray}
we can write Eqs.(\ref{eq:f-var2}) and (\ref{eq:f-var-trace}) as
\begin{equation}\label{eq:f-var-st}
G_{\mu \nu }=\frac{\kappa }{\phi}{T}_{\mu \nu }-\frac{V(\phi)}{2\phi}{g}_{\mu \nu }+\frac{1}{\phi}\left(\phi_{;\
\mu\nu}-{g}_{\mu \nu }{\Box}\phi\right)
\end{equation}
\begin{equation}\label{eq:f-var-st2}
3\Box \phi + 2V(\phi)-\phi\frac{dV}{d\phi}=\kappa T \label{eq:Box-f}\,,
\end{equation}
 which can also be obtained from a Brans-Dicke action of the form
\begin{equation} \label{eq:ST}
{\cal A}_{\phi}=\int d^4 x\sqrt{-{g}}\left[\phi {R}-V(\phi) +\kappa {\cal L}_m\right]\,.
\end{equation}
This expression is related to the so called {\it O'Hanlon Lagrangian}, which belongs to a class of Lagrangians
introduced in order to
achieve a covariant model for a massive dilaton theory \cite{o'hanlon}.\\
It is evident that the Lagrangian (\ref{eq:ST}) is very similar to a Brans-Dicke theory, but is lacking of the
kinetic term. The formal analogy between the Brans-Dicke scheme and fourth order gravity schemes is obtained in
the particular case $\omega_{BD} =0$.\\
If we consider the matter term vanishing, Eq.(\ref{eq:f-var-trace}) becomes an ordinary Klein-Gordon equation,
where $f'(R)$ plays the role of an effective scalar field whose mass is determined by the self-interaction
potential.\\

\section{PPN-Formalism in Scalar Tensor Gravity}

Along this paper, we base our discussion on the analogy between scalar-tensor theories of gravity and the higher
order ones to analyze the problem of the PPN-limit for the fourth order gravity model. Recently the cosmological
relevance of higher order gravity has been widely demonstrated. On the other side, the low energy limit of such
theories is still not satisfactory investigated, although some results on the galactic scales have been already
achieved \cite{newtlim}. A fundamental test to understand the relevance of such a scheme is to check if there is
even an accord with Solar System experiments. As outlined in the introduction, some controversial results have
been recently proposed \cite{olmo,allemandi-ruggiero}. To better develop this analysis, we can refer again to the
scalar-tensor\,-\,higher order gravity analogy, exploiting the PPN results obtained in the scalar-tensor scheme \cite{esposito-farese}.\\
A satisfactory description of PPN limit for this kind of theories has been developed in
\cite{esposito-farese,damour1,damour2}. In these works, the problem has been treated providing interesting results
even in the case of strong gravitational sources like pulsars and neutron stars where the deviations from General
Relativity are obtained in
 a non-perturbative regime \cite{damour2}.
A clear summary of this formalism can be found in the papers \cite{esposito-farese} and \cite{schimd05}.\\
The action to describe a scalar-tensor theory can be assumed, in natural units, of the form (\ref{L-scal-tens}).
The matter Lagrangian density is again considered depending only on the metric $g_{\mu\nu}$ and the matter fields.
This action can be easily redefined in term of a minimally coupled scalar field model {\it via} a conformal
transformation of the form $g^*_{\mu\nu}\,=\,F(\phi)g_{\mu\nu}$. In fact, assuming the transformation rules:
\begin{equation}\label{conf-phi}
\left(\frac{d\psi}{d\phi}\right)^2\,=\,\frac{3}{4}\left(\frac{d\ln{F(\phi)}}{d\phi}\right)^2+\frac{1}{2F(\phi)}\,,
\end{equation}
and
\begin{equation}
A(\psi)\,=\,F^{-1/2}(\phi)\,,\ \ \ \ \ \ \ \ V(\psi)\,=\,V(\phi)F^{-2}(\phi)\,,
\end{equation}
\begin{equation}
{\cal L}_m^*\,=\,{\cal L}_m\,F^{-2}(\phi)\,,
\end{equation}
one gets the action
\begin{equation}\label{scatenLag*}
{\cal A}_{*}\,=\,\int{\sqrt{-g_*}\left[R_*+\frac{1}{2}g_{*}^{\mu\nu}\psi_{,\mu}\psi_{,\nu}-V(\psi)+{\cal
L^*}_{m}\right]}\,.
\end{equation}
The first consequence of such a transformation is that now the non-minimal coupling is transferred on the ordinary
matter sector. In fact, the Lagrangian ${\cal L}^*_m$ is dependent not only on the conformally transformed metric
$g^*_{\mu\nu}$ and the matter field but it is even characterized by the coupling function $A(\psi)^2$. In the same
way, the field equations can be recast in the Einstein frame. The energy-momentum tensor is defined as
$T^{m\,*}_{\mu\nu}\,=\,\frac{2}{\sqrt{-g^*}}\frac{\delta {\cal L}_m}{\delta{g^*_{\mu\nu}}}$ and it is related to
the Jordan expression as $T^{m\,*}_{\mu\nu}\,=\,A(\psi)T^{m}_{\mu\nu}$. The function:
\begin{equation}\label{alpha}
\alpha (\psi)\,=\,\frac{d \ln{A(\psi)}}{d\psi}
\end{equation}
establishes a measure of the coupling arising in the Einstein frame between the scalar sector and the matter one
as an effect of the conformal transformation (General Relativity is recovered when this quantity vanishes). It is
possible even to define a control of the variation of the coupling function through the definition of the
parameter $\beta\,=\,\displaystyle\frac{d {\alpha(\psi)}}{d\psi}$. Regarding the effective gravitational constant,
it can be expressed in term of the function $A(\psi)$ as $G_{eff}\,=\,\frac{G_N}{F(\phi)}\,=\,G_N\,A^2(\psi)$. It
has to be remarked that such a quantity is, in reality, well different by the Newton constant measured in the
Cavendish-like terrestrial experiments (see Eq.(\ref{cavendish-sc}) below).\\
Let us now, concentrate on the scalar-tensor generalization of the local gravitational constraints. Deviations
from General Relativity can be characterized through Solar System experiments \cite{will} and binary pulsar
observations which give an experimental estimate of the PPN parameters. These parameters were introduced by
Eddington to better determine the deviation from the standard prediction of General Relativity,
expanding local metrics as the Schwarzschild one, to higher order terms.\\
The generalization of this quantities to scalar-tensor theories allows the PPN-parameters to be expressed in term
of non-minimal coupling function $F(\phi)$ or, equivalently, in term of the parameter $\alpha$ defined in
Eq.(\ref{alpha}), that is\,:
\begin{equation}\label{gamma}
\gamma^{PPN}-1\,=\,-\frac{(F'(\phi))^2}{F(\phi)+2[F'(\phi)]^2}\,=\,-2\frac{\alpha^2}{1+\alpha^2}\,,
\end{equation}
\begin{displaymath}
\beta^{PPN}-1\,=\,\frac{1}{4}\frac{F(\phi)\cdot F'(\phi)}{2F(\phi)+3[F'(\phi)]^2}\frac{d\gamma^{PPN}}{d\phi}\,=
\end{displaymath}
\begin{equation}\label{beta}
=\,\frac{1}{2}\frac{\alpha^2}{(1+\alpha^2)^2}\frac{d\alpha}{d\psi}\,.
\end{equation}
The above definitions imply that the PPN-parameters become dependent on the non-minimal coupling function
$F(\phi)$ and its derivatives. They can be directly constrained by the observational data. Actually, Solar System
experiments give accurate indications on the ranges of $\gamma^{PPN}_0\,,\ \beta^{PPN}_0$\footnote{We indicate
with the subscript $_0$ the Solar System measured estimates.}. Results are summarized in Tab.\ref{ppn}.
\begin{table}[h]
\centering
\begin{tabular}{|l|c|}
\hline\hline
  Mercury Perih. Shift& $|2\gamma_{0}^{PPN}-\beta_{0}^{PPN}-1|<3\times10^{-3}$ \\\hline
 Lunar Laser Rang. &  $4\beta_{0}^{PPN}-\gamma_{0}^{PPN}-3\,=\,-(0.7\pm 1)\times{10^{-3}}$ \\\hline
 Very Long Bas. Int.  &  $|\gamma_{0}^{PPN}-1|\,=\,4\times10^{-4}$ \\\hline
 Cassini spacecraft &  $\gamma_{0}^{PPN}-1\,=\,(2.1\pm 2.3)\times10^{-5}$ \\\hline\hline
\end{tabular}
\caption{\small \label{ppn} A schematic resume of recent constraints on the PPN-parameters. They are the
perihelion shift of Mercury \cite{mercury}, the Lunar Laser Ranging \cite{lls}, the upper limit coming from the
Very Long Baseline Interferometry \cite{VLBI} and the results obtained by the estimate of the Cassini spacecraft
delay into the radio waves transmission near the Solar conjunction \cite{cassini}.}
\end{table}
The experimental results can be substantially resumed into the two limits \cite{schimd05}\,:
\begin{equation}
|\gamma^{PPN}_0-1|\leq{2\times10^{-3}}\,,\ \ \ \ \ |\beta_0^{PPN}-1|\leq{6\times{10^{-4}}}\,,
\end{equation}
which can be converted into constraints on $\alpha_0$ and $\beta_0$. In particular, the Cassini spacecraft value
induces the bound $\alpha_0\sim\displaystyle{\frac{F_{0\ ,\phi}}{F_0}}\,<\,4\times 10^{-4}$. At first sight, one
can deduce that the first derivative of the coupling function $A(\psi)$ has to be very small, which means a very
low interaction between matter and the scalar field; conversely the second derivative $\beta_0$ can take large
values so that the matter
sector may be strongly coupled with scalar degrees of freedom \cite{esposito-farese}.\\
Together with the Solar System experiments, even binary-pulsar tests can be physically significant to characterize
the PPN-parameters. From this analysis \cite{damour1,damour2,esposito-farese} descends that the
second derivative can be a large number, i.e. $\beta_0\,>\,-4.5$ even for a vanishingly small $\alpha_0$.\\
This constraint allows to achieve a further limit on the two PPN-parameters $\gamma^{PPN}$ and $\beta^{PPN}$,
which can be outlined by means of the ratio\,:
\begin{equation}\label{pulsar}
\frac{\beta ^{PPN}-1}{\gamma^{PPN}-1}\,<\,1.1\,.
\end{equation}
The singular $(0/0)$ nature of this ratio puts in evidence that it was not possible to get such a limit in the
case of weak\,-\,field experiments (see for details \cite{esposito-farese}).\\
For sake of completeness, we cite here even the shift that the scalar-tensor gravity induces on the theoretical
predictions for the local value of the gravitational constant as coming from Cavendish-like experiments. This
quantity represents the gravitational coupling measured when the Newton force arises between two masses\,:
\begin{equation}\label{cavendish} G_{Cav}\,=\,\frac{F\cdot r^2}{m_1\cdot m_2}\,.
\end{equation}
In the case of scalar tensor gravity, the Cavendish coupling reads\,:
\begin{displaymath}
G_{Cav}\,=\,\frac{G_N}{F(\phi)}\left[1+\frac{[F'(\phi)]^2}{2F(\phi)+3[F'(\phi)]^2}\right]\,=
\end{displaymath}
\begin{equation}\label{cavendish-sc}
=\,G_N\,\,A^2(\psi)(1+\alpha^2)\,.
\end{equation}
From the limit on $\alpha$ coming from Cassini spacecraft, the difference between $G_{Cav}$ and $G_{eff}$ is not
more than the $10^{-3}\%$.

\section{PPN limit of fourth order gravity inspired by the scalar-tensor analogy}

In previous section, we discussed the PPN limit in the case of a scalar-tensor gravity. These results can be
extended to the case of fourth order exploiting the analogy with scalar-tensor case developed in Sec. 2.\\
We have seen that fourth order gravity is equivalent to the introduction of a scalar extra degree of freedom into
the dynamics. In particular, from this transformation, it derives a Brans-Dicke type Lagrangian with a vanishing
Brans-Dicke parameter $\omega_{BD}\,=\,0$. Performing the change of variables implied by a conformal
transformation, the Brans-Dicke Lagrangian can be furtherly transformed into a Lagrangian where the non-minimal
coupling is moved onto the matter side as in (\ref{scatenLag*}). The net effect is that, as in the case of a
``true" scalar-tensor theory, it is possible to develop an Einstein frame formalism which allows a PPN-limit
analysis. The basic physical difference between the two descriptions is that the quantities entering the
PPN-parameters $\gamma^{PPN}$ and $\beta^{PPN}$, or the derivatives of the non-minimal coupling function
$A(\psi)$, are now $f'(R)$ and its derivatives with respect to the Ricci
scalar $R$ since the non minimal coupling function in the Jordan frame is $f'(R)\equiv \phi$.\\
Alternatively, to obtain a more versatile equivalence between the two approaches it is possible to write down
fourth order gravity by an analytic function of the Ricci scalar considering the identification induced by the
field equations, i.e. $\varphi\rightarrow R$. In fact, if one takes into account the scalar-tensor Lagrangian\,:
\begin{equation}\label{scaten-Rphi}
\int{d^4\,\sqrt{-g}\left[ \,F(\varphi)+(R-\varphi)F'(\varphi)+\kappa{\cal L}_m\right]}\,,
\end{equation}
the variation with respect $\varphi$ and the metric provide the above identification and a system of field
equations which are completely equivalent to the ordinary ones descending from fourth order gravity. The
expression (\ref{scaten-Rphi}) can be recast in the form of the O'Hanlon Lagrangian (\ref{eq:ST}) by means of the
substitutions\,:
\begin{equation}\label{ohanl-scaten}
\phi\equiv\,F'(\varphi)\,,\ \ \ \ \ \ \ \ \ \ V(\phi)\equiv\,\varphi F'(\varphi)-F(\varphi)\,,
\end{equation}
where, in such a case, the prime means the derivative with respect to $\varphi$. It is evident that the new
scalar-tensor description implies a non-minimal coupling function through the term
\begin{equation}\label{F'(phi)-F'(R)}
F'(\varphi)\,=\,\frac{df(R)}{dR}\,,
\end{equation}
and the identification $\varphi\rightarrow R$ implies that the higher order derivatives can be straightforwardly
generalized.\\
At this point, it is immediate to extend the results of the PPN-formalism developed for scalar-tensor gravity to
the case of a fourth order theory. In fact, it is possible to recast the PPN parameters (\ref{gamma})-(\ref{beta})
in term of
the curvature invariants quantities.\\
This means that the non-minimal coupling function role, in the fourth order scenario, is played by the $df(R)/dR$
quantity. As a consequence the PPN-parameters (\ref{gamma}) and (\ref{beta}) become\,:
\begin{equation}\label{gammaR}
\gamma^{PPN}_R-1\,=-\,\frac{f''(R)^2}{f'(R)+2f''(R)^2}\,,
\end{equation}
\begin{equation}\label{betaR}
\beta^{PPN}_R-1\,=\,\frac{1}{4}\frac{f'(R)\cdot f''(R)}{2 f'(R)+3f''(R)^2}\frac{d\gamma^{PPN}_R}{d\phi}\,.
\end{equation}
These quantities have, now, to fulfill the requirements drawn from the experimental tests resumed in Table
\ref{ppn}. The immediate consequence of such definitions is that derivatives of fourth order gravity theories have
to satisfy constraints in relation to the actual measured values of the Ricci scalar $R_0$. As a matter of fact,
one can check these quantities by the Solar System experimental prescriptions and deduce the compatibility between
fourth order gravity and General Relativity.\\
Since the definitions (\ref{gammaR}) and (\ref{betaR}) do not allow to obtain, in general, upper limits on $f(R)$
from the constraints of Table 1, one can arbitrarily chose classes of fourth order Lagrangians, in order to check
if the approach is working. We shall adopt classes of Lagrangians which are interesting from a cosmological point
of view since give viable results to solve the dark energy problem \cite{curv-quint,noi-review,noi-ijmpd}.
\\
In principle, one can try to obtain some hints on the form of $F(\varphi)$ (or correspondently of the $f(R)$) by
imposing constraints provided from the Lunar Laser Ranging (LLR) experiments and the Cassini spacecraft
measurements which give direct stringent estimates of PPN-parameters. After, one can try to solve these relations
and then to verify what is the response to the pulsar upper limit with respect to the ratio
$\displaystyle{\frac{\beta^{PPN}-1}{\gamma^{PPN}-1}}\,<\,1.1$. This procedure has shown that generally if the two
Solar System relations are verified, the pulsar constraint is well fitted by a modified gravity model. However
this result is strictly influenced by the error range of the Cassini and LLR tests.\\
After this remark, one can consider different fourth order Lagrangians with respect to the two Solar System
constraints coming from the perihelion shift of Mercury and the Very Long Baseline Interferometry.\\
The results are summarized in Table \ref{table2}. We have listed the fourth order Lagrangians considered in the
first column and the limit on the model parameters induced by the Solar System constraints is in the second
column.
\begin{table}[h]
\centering
\begin{tabular}{|c|c|}
\hline\hline Lagrangian & Parameters constraints
\\ \hline
$f_0 R^2$ &\\& $R_0\,<\,0,\ \frac{R_0}{4996}\,<\,f_0\,<\,-\frac{R_0}{5004}$\\&\\& $R_0\,>\,0,\
-\frac{R_0}{5004}\,<\,f_0\,<\,\frac{R_0}{4996}$\\&\\\hline $f_0\,R^3$ &\\&
$-\frac{1}{30024}\,<\,f_0\,<\,\frac{1}{29976}$\\&
\\ \hline
 $R\,+\,aR^2$& $a\,=\,0$ \\&\\& $\left\{a\,>\,0\,,\ 9992a\,<\,\frac{1}{a}+2R_0\right\}$\\&\\&
 $\left\{a\,>\,0\,,\ \frac{1}{a}+10008a +2R_0\,<\,0\right\}$ \\&\\&
 $\left\{a\,<\,0\,,\ \frac{1}{a}+10008a +2R_0\,>\,0\right\}$ \\&\\&
 $\left\{a\,<\,0\,,\ 9992a\,>\,\frac{1}{a}+2R_0\right\}$\\\hline
$A\,\log[R]$& $A\,\leq\,0\,,\ R_0\,<\,13.5685 A^{1/3}$ \\&\\& $R_0\,>\,-13.5757 A^{1/3}$ \\&\\& $A\,>\,0\,,\
R_0\,<\,-13.5757 A^{1/3}$ \\&\\& $R_0\,>\,13.5685 A^{1/3}$
 \\\hline\hline
\end{tabular}
\caption{\small \label{table2} Constraints induced by PPN-experimental upper bounds for different cases of fourth
order gravity Lagrangians. Solar System experiments are the Mercury Perihelion Shift and the Very Long Baseline
Interferometry.}
\end{table}
As it is possible to see, the PPN-limits induced by the Solar System tests can be fulfilled by different kinds of
fourth order Lagrangians provided that their parameters remain well defined with respect to the
background value of the curvature.\\
These results corroborate evidences for a defined PPN-limit which does not exclude higher order gravity. They are
 in contrast with other recent investigations \cite{olmo,ppn-bad}, where it has been pointed out
that this kind of theories are not excluded by experimental results in the weak field limit and with respect to
the PPN prescriptions.\\
Similar results also hold for Lagrangians as $f(R)\,=\,f_0R^n$ and $f(R)\,=\,R\,+\,\frac{\mu}{R}$ which have shown
 interesting properties from a cosmological point of view \cite{curv-quint,noi-review,noi-ijmpd,carroll}. This
fact allows to establish a significant link between gravity at local and cosmological scales.\\
Finally a remark is in order. It has to be taken into account that the $f(R)\,=\,A\ln[R]$ does not admit a
Minkowski background around which to perform the usual post-Newtonian analysis. Due to this fact this model is
essentially different from the others in the weak energy limit.

\section{Conclusions}

Since the issue of higher order gravity is recently become a very debated matter, we have discussed its low energy
limit considering the PPN-formalism in the metric framework. The study is based on the analogy between the
scalar-tensor gravity and fourth order gravity. Such an investigation is particularly interesting even in relation
to the debate about the real meaning of the curvature fluid which could be a natural explanation for dark energy
\cite{curv-quint,noi-review,noi-ijmpd,carroll,odintsov-m,palatini,francaviglia,olmo,allemandi-ruggiero}. The
PPN-limit indicates that several fourth order Lagrangians could be viable on the Solar System scales. It has to
remarked that the Solar System experiments pose rather tight constraints on the values of coupling constants, e.g.
$f_0$ (see Table II). Such a result does not agree with the very recent papers \cite{olmo} which suggest negative
conclusions in this sense, based on questionable
 theoretical assumptions and extrapolations.\\
 It is evident that such discussion does not represent a final
 answer on this puzzling issue. Nevertheless it is reasonable to
 affirm that extended gravity theories cannot be ruled out, definitively, by Solar
 System experiments. Of course, further accurate investigations are needed to
 achieve some other significant indications in this sense, both
 from theoretical and experimental points of view. For example
 the study of higher order gravity PPN-limit directly in the
 Jordan frame could represent an interesting task for
 forthcoming investigations.\\
An important concluding remark is due at this point. A scalar-tensor theory can be recast in the Einstein frame,
{\it via} a conformal transformation, implying an equivalent framework. Actually, dealing with higher order
gravity, there is no more such a conformal transformation able to ``equivalently" transform the whole system from
the Jordan frame to the Einstein one. Effectively, it is possible to conformally transform a higher order (and, in
particular, a fourth order) theory into an Einstein-like with the addiction of some scalar fields as a direct
consequence of the equivalence between the higher order framework and the scalar-tensor one at level of the
classical field equations. This equivalence addressed, as {\it dynamical equivalence} \cite{wands:cqg94}, does not
holds anymore when one considers configurations which do not follow the classical trajectories, for example in the
case of quantum effects. A fundamental result which follows from this considerations is that dealing with the
early-time inflationary scenario one can safely perform calculations for the primordial perturbations in the
Einstein conformal frame of a scalar-tensor model while it is not possible to develop such calculations in the
case of an higher order gravity scenario since the scalar degrees of freedom are no more independent of the
gravitational field source. This issue holds, if the effective field is induced from geometrical degrees of
freedom. Since the PPN-limit is achieved in the semi-classical limit, when the conformal factor turns out to be
well defined, deductions about the PPN-limit for fourth order gravity models, developed exploiting the analogy
with the scalar-tensor scheme, are safe from problems.

\end{document}